\documentclass[a4paper,11pt]{article}
\usepackage{pos}
\usepackage{physics}
\usepackage{aas_macros}
\title{A Hierarchical Framework for explaining the Cosmic Ray Spectrum using Diffusive Shock Acceleration}
 \ShortTitle{Hierarchical Framework for Cosmic Ray Acceleration}
\author*[a]{Roger Blandford}
\author[b]{Paul Simeon}
\author[c,d]{Noémie Globus}
\author[e]{Payel Mukhopadhyay}
\author[f]{Enrico Peretti}
\author[g]{Kirk S. S. Barrow}
\affiliation[a]{KIPAC, Stanford University, 452 Lomita Mall, Stanford, CA 94305, USA}
\affiliation[b]{Independent, Los Altos, CA, USA}
\affiliation[c]{Department of Astronomy and Astrophysics, University of California, Santa Cruz, CA 95064, USA}
\affiliation[d]{Astrophysical Big Bang Laboratory, RIKEN, Wako, Saitama, Japan}

\affiliation[e]{Department of Physics, University Ave., Oxford St., Berkeley, CA94720, USA}
\affiliation[f]{Niels Bohr International Academy, Niels Bohr Institute, University of Copenhagen, Blegdamsvej 17, DK-2100 Copenhagen, Denmark}
\affiliation[g]{Astronomy Department, University of Illinois at Urbana-Champaign, 1002 W Green St, Urbana, IL 61801, USA}
\emailAdd{rdb3@stanford.edu}
\emailAdd{paulsimeon@gmail.com}
\emailAdd{noglobus@ucsc.edu}
\emailAdd{pmukho@berkeley.edu}
\emailAdd{enrico.peretti@gssi.it}
\emailAdd{kbarrow@illinois.edu}
\abstract{The hypothesis that the entire cosmic ray spectrum, from $\lesssim1\,{\rm GeV}$ to $\gtrsim100\,{\rm EeV}$ energy, can be accounted for by diffusive shock acceleration on increasingly large scales is critically examined. Specifically, it is conjectured that Galactic cosmic rays, up to $\sim3\,{\rm PeV}$, are mostly produced by local supernova remnants, from which they escape upstream. These cosmic rays initiate a powerful magnetocentrifugal wind, removing disk mass and angular momentum before passing through the Galactic Wind Termination Shock at a radius $\sim200\,{\rm kpc}$, where they can be re-accelerated to account for observed cosmic rays up to $\sim30\,{\rm PeV}$. The cosmic rays transmitted downstream from more powerful termination shocks associated with other galaxies can be further accelerated at Intergalactic Accretion Shocks to the highest energies observed. In this interpretation, the highest rigidity observed particles are protons; the highest energy particles are heavy nuclei, such as iron. A universal "bootstrap" prescription, coupling the energy density of the magnetic turbulence to that of the resonant cosmic rays, is proposed, initially for the highest energy particles escaping far ahead of the shock front and then scattering, successively, lower energy particles downstream. Observable implications of this general scheme relate to the spectrum, composition and sky distribution of Ultra-High-Energy Cosmic Rays, the extragalactic radio background, the Galactic halo magnetic field and Pevatrons.}
\ConferenceLogo{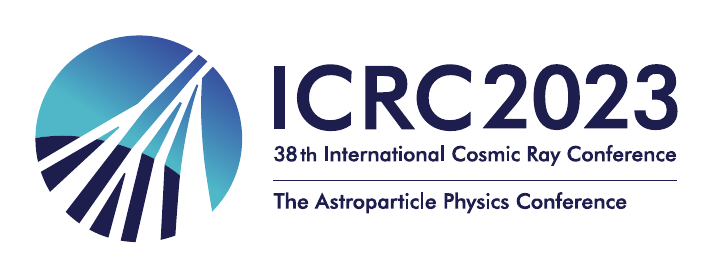}
\FullConference{38th International Cosmic Ray Conference (ICRC2023)\\26 July - 3 August, 2023\\Nagoya, Japan}
\begin{document}
\maketitle
\section{Hierarchical Cosmic Ray Acceleration}\label{sec:HCRA}
Cosmic rays, our first non-electromagnetic astronomical messenger, were discovered over a century ago. It is generally accepted that most GeV--TeV particles are accelerated at SuperNova Remnants, SNR. It is widely agreed that they are produced by Diffusive Shock Acceleration, DSA \cite[e.g.,][]{drury12}, which can account for the energetics and the observed power-law distribution, after allowing for energy-dependent propagation. This spectrum steepens above $\sim3\,{\rm PeV}$, and several possible sources, including pulsars and their nebulae, the Galactic Center and the Galactic Wind Termination Shock, GWTS \cite[][and referencs therein]{mukhopadhyay23}, have been proposed. The highest energy cosmic rays ($E\gtrsim1\,{\rm EeV}$) are generally argued to be extragalactic in origin, and proposed sources include relativistic jets from stellar and supermassive black holes, and large-scale, nonrelativistic Intergalactic Accretion Shocks, IAS \citep{norman95,Simeon:2023UF}.
\begin{figure}[h]
\centering
\includegraphics[width=1\textwidth]{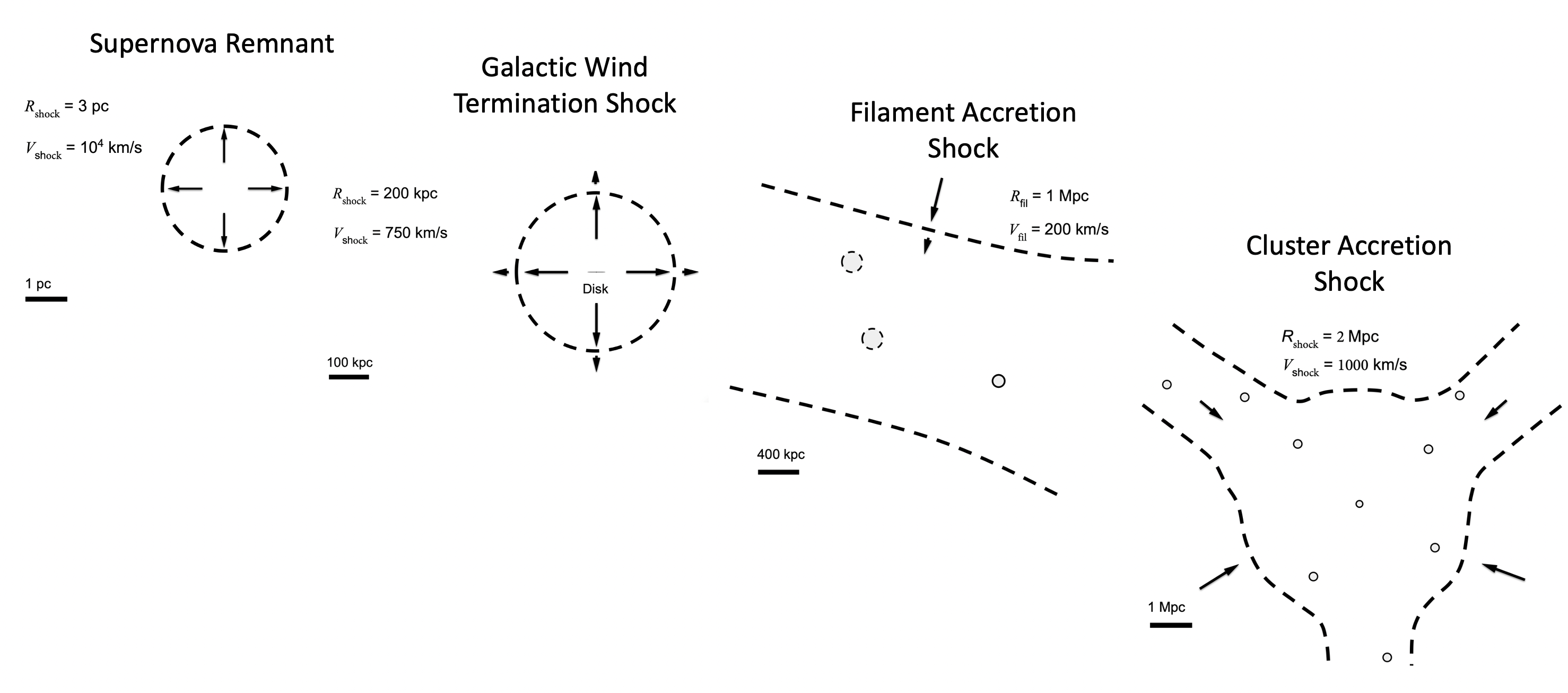}
\caption{Montage displaying major features of the hierarchical framework for explaining cosmic ray acceleration from $\lesssim1\,{\rm GeV}$ to $\gtrsim100\,{\rm EeV}$. Non-relativistic cosmic rays in the hot interstellar medium are accelerated by supernova shock waves up to $\sim$ PeV energies, with the most energetic particles escaping ahead of the shock front to become most of the observed Galactic cosmic ray distribution. These cosmic rays drive an MHD wind through the Galactic halo eventually passing through a termination shock where more energetic particles are accelerated. Those that escape upstream contribute to the observed $\gtrsim\,{\rm PeV}$ spectrum; those that are transmitted downstream join outflows from more powerful galaxies to form the intergalactic cosmic ray distribution. Intergalactic cosmic rays can fall into strong intergalactic shock fronts, notably those surrounding rich clusters of galaxies where they can be re-accelerated to EeV energies.}
\label{fig:montage}
\end{figure}

In this report, we describe one approach to explaining the entire cosmic ray spectrum using one mechanism --- DSA --- in a variety of locales (Fig.~\ref{fig:montage}). We do this less in the spirit of strong advocacy and more with the wish to exploit a common physical description of DSA to derive observable implications which can falsify the general model. To be specific, we postulate that suprathermal particles from stellar winds are accelerated by expanding supernova shock fronts and that the observed GeV--PeV spectrum combines the highest energy particles, which escape upstream, with the lower energy particles, which are transmitted downstream.

All of these cosmic rays must escape the Galactic disk in $\lesssim10\,{\rm Myr}$ as a hydromagnetic wind carrying off mass, angular momentum and energy from the Galactic disk. This wind passes through a termination shock at the periphery of the (dark matter-dominated) Galaxy at $\sim200\,{\rm kpc}$, where cosmic rays are further re-accelerated. Those that escape upstream into the halo and the disk contribute to the ``shin'' of the observed spectrum; those that are transmitted downstream join even more powerful outflows from active, including starburst, galaxies to build up the PeV cosmic ray population in the intergalactic medium, evolving to form a ``cosmic web,'' of IAS including quasi-spherical shocks around rich clusters, quasi-cylindrical shocks around filaments, and sheets. We may live in a filament or sheet which may contribute cosmic rays with energies up to $\sim3\,{\rm EeV}$ transmitted downstream and contributing to the spectrum we observe. Finally, we propose that the highest energy cosmic rays --- up to $\sim200\,{\rm EeV}$ --- derive from upstream escape of particles accelerated at the strongest, nearby, intergalactic shocks like the accretion shocks surrounding the Virgo Cluster (17 Mpc) and more distant, stronger shocks surrounding richer clusters. The sources must be relatively nearby because these extreme energy cosmic rays have comparatively short lifetimes in the cosmic microwave background \cite[e.g.,][and Globus et al., these proceedings]{Globus:2023EA}. 

\section{Generic Diffusive Shock Acceleration}\label{sec:DSA}
The central idea behind DSA is that high energy particles are scattered by a spectrum of hydromagnetic waves so that they diffuse with coefficient $D$ relative to the background plasma moving with velocity $\bf u$. Their energies change little so long as $\bf u$ changes slowly. However, at a shock front, the speed of the plasma changes abruptly and cosmic rays experience a relative gain, $O(u/c)$, measured in a frame moving with the gas. A typical cosmic ray crosses the shock $O(c/u)$ times leading to an energy gain $O(1)$. This is a statistical (Fermi-like) process, and an exponentially small number of particles will have an exponentially large gain in energy leading to a power-law spectrum. The spectrum of the accelerated particles depends upon the geometry of the flow and the microphysics of the shock. It is important to allow for the divergence/convergence of the flow, which affects the particle acceleration and escape. 

A common feature of DSA is that it must be near maximally efficient, at least when the shock is strong if it is to account for the full spectrum of cosmic rays. This has the immediate implication that the cosmic rays participate in the dynamics of the flow and change both $\bf u$ and $D$. Most discussions of DSA have either assumed or attempted to calculate a form for $D$. Sometimes these have been inspired by linear or quasilinear growth of plasma instabilities; sometimes, the evolution from a non-accelerating initial state is followed using a Particle-In-Cell, PIC, code. These simulations have taught us much but lack the dynamic range and geometry needed to capture the entire problem. Mostly, they have not demonstrated the high levels of scattering required for maximal acceleration. In this report, we ask what form of $D$ might account for the observations and then ask under what circumstances could it be sustained.

Let us first set up a formalism appropriate to an idealized model of a quasi-spherical IAS surrounding a rich cluster but adaptable to other shocks. Specifically, we consider a stationary spherical gas inflow with speed $u(r)$ towards a spherical shock with radius $r_{\rm shock}$ and speed $u_{\rm shock}$. The corresponding gas density is $\rho_{\rm shock}$, and the rate of flow of kinetic energy across the shock is $L_{\rm shock}=2\pi r_{\rm shock}^2\rho_{\rm shock}u_{\rm shock}^3$. We consider protons with rigidity $R$, which we measure in units of a rigidity scale $R_0$ and introduce $q\equiv\ln(R/R_0)$. We posit the presence of a spatial diffusion coefficient $D(r,q)$. The usual equation for cosmic ray transport ahead of a strong, nonrelativistic shock front describes the evolution of the isotropic part of the particle distribution function in the frame of the gas, $f(r,R)$, under a combination of convection by the gas and diffusion through it. It is convenient to replace $f$ with $N(r,q)=16\pi^2r^2R^3f$, which is the number of cosmic rays per unit $r,q$. The equation of cosmic ray particle conservation can then be written as 
\begin{equation}\label{eq:cons}
\partial_t N+\partial_r[-Nu-D\left(\partial_r N-2N/r\right)]+\partial_q[N\dot q]=0,
\end{equation}
where $\dot q\equiv dq/dt$ combines the adiabatic acceleration due to compression of the gas, $\dot q_{\rm ad}=(1/3r^2)d(r^2u)/dr$, and radiative loss $\dot q_{\rm loss}$ due to interaction with the microwave background.  We can regard the two expressions within square brackets as components of particle flow vector ${\bf F}=\{F_r,F_q\}$.

Formally, the rate of cosmic ray acceleration is determined at, and in the frame of, the shock front by imposing continuity of the particle flux at a given rigidity. We optimize the acceleration by supposing that the shock is strong with compression ratio $4$ and that there is no diffusion downstream of the shock. This implies that $F_r=-(\partial_qN+N)/4$ at the shock. For small enough values of $q$, we suppose that protons in the intergalactic medium are swept into the shock front, where they are subject to DSA.  $F_r$ is negative, the diffusion scale height $\sim D/u$ is much smaller than $r$ so the shock is effectively planar. We recover the usual Green's function solution, $N(1,q)=-4e^{-q}\int_0^qdq'e^{q'}F_r(1,q')$, so that the slope of the spectrum $N(1,q)$ falls off more slowly than $e^{-q}$. 

We next suppose that there is some intermediate range of $q$ for which the supply of protons ahead of the shock can be ignored. Particles of lower $q$ will be accelerated with a spectrum $N(1,q)\propto e^{-q}$, the scale height will still be $\ll r_{\rm shock}$, and there will be an absence of particles with this $q$ far upstream. In this regime, $F_r\sim0$. However, the scale height will increase with $q$ and eventually become $\sim r$ so that the protons can escape with $F_r>0$. It is helpful to ignore the downstream flow and introduce a fictitious surface current along $q$ at the shock so that inflowing particles at low $q$ become escaping particles at high $q$. Note that if the mean free path $\ell$ satisfies $ur/c\lesssim\ell\lesssim r$, the particle can escape and still propagate diffusively.

\section{Bootstrap Process}\label{sec:bootstrap}
Efficient acceleration depends upon maintaining a short scattering mean free path. Most analyses to date, both analytical and computational, have focused on the growth of various instabilities close to the subshock.  However, efficient acceleration to $\gtrsim\,{\rm EV}$ rigidity at an IAS requires that $\lesssim\,{\rm nG}$ magnetic field in the IGM become $\gtrsim\,\mu{\rm G}$ field, and we suppose that this grows spatially towards the shock through interaction with particles that escape upstream and impedes their escape. This "bootstrap" \citep{blandford07} process initiates magnetic field growth in a manner that will be idiosyncratic to the conditions and will involve a combination of MHD, resonant and ``Bell'' instabilities \citep[e.g.,][]{schroer22}. The field will quickly become nonlinear and strongly turbulent with no preferred direction and spanning a range wavevectors, $k$. At this point we suppose that the primary interaction involves particles of rigidity $R$ and Larmor radius $r_g\sim k^{-1}\sim R/B(R)$ (implicitly and  in SI units) interacting with waves of spectral energy density, per $q$, $W_{{\rm mag}\,q}\sim B(R)^2/\mu_0$. We then suppose that the mean free path is $\ell(r,q)\sim r_g$. (This is a spectral generalization of the ``Bohm'' hypothesis.) 

Now $W_{{\rm mag}\,q}$ will be determined by the resonant particles with corresponding spectral energy density $W_{{\rm part}\,q}=RN(r,q)/4\pi r^2$. We next introduce a simple, equipartition ansatz, that $W_{{\rm mag}\,q}\sim W_{{\rm part}\,q}$, again eschewing constants $O(1)$. Adopting these relations, we obtain an expression for the nonlinear diffusion coefficent $D(r,q)=\ell(r,q)c/3\sim(rc/3)(4\pi R/\mu_0Nec)^{1/2}$. The particle - wave interaction time will be of order the wave period which is of order the time for wave energy to flow through $k$-space and shorter than the flow time, $\sim(c/u)\ell$.

There are two relevant limits to the maximum rigidity of the particles that escape upstream, $R_{\rm max}$. The first, $R_{\rm diff}$, comes about because the energy density of the resonant waves at $R_{\rm max}$ at a distance $\sim r_{\rm shock}$ ahead of the shock front should be at most a modest fraction of the gas kinetic energy density $\sim\rho u^2/2$. Otherwise it and the associated cosmic-ray energy will decelerate the gas and weaken the shock. This sets a limit $R_{\rm diff}\sim(\mu_0L_{\rm shock}u/4\pi c^2)^{1/2}$. This bound is related to, though ultimately lower than that given by \cite{hillas84}. Our prescription probably maximizes $R_{\rm diff}$ at a given shock. A more complete, time-dependent treatment, will include mass, momentum and energy conservation at the fluid level. The second limit, $R_{\rm loss}$, arises from setting the radiative loss time $|{\dot q}_{\rm loss}|^{-1}$ to the acceleration/flow time $\sim r_{\rm shock}/u_{\rm shock}$.

\begin{figure}[h]
\centering
\includegraphics[width=1\textwidth]{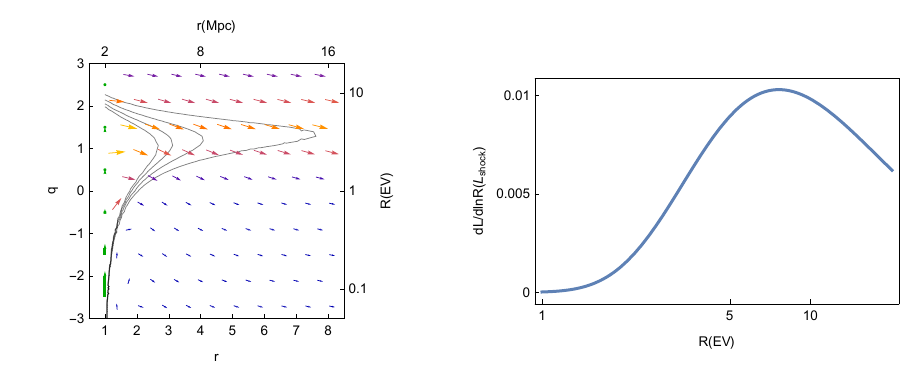}
\caption{(a) Contours of proton number, (per $r$, $q$), $N(r,q)$, in $r-q$ space ahead of a strong cluster accretion shock modeled on  the Virgo cluster. There is supposed to be a source of intergalactic cosmic rays with $q< -3$ that sources a shock surface current flowing towards positive $q$, shown as green arrows at $r=1$. This surface current falls slowly for $-3\lesssim q\lesssim0$ as the upstream cosmic rays have scale-heights smaller than $r$ and essentially no particles escape upstream. For $0\lesssim q\lesssim3$, an increasing fraction of the particles escape while experiencing modest adiabatic acceleration and increasingly strong radiative loss. The arrows show the flow of protons in $r-q$ space. (b) Spectral proton luminosity, $dL/dq$ at the Galaxy. The peak of the spectrum of  escaping protons in this example has a rigidity $\sim7\,{\rm EV}$ and an energy $\sim7\,{\rm EeV}$. Heavy nuclei, such as iron, propagate in the same fashion at the same rigidity but are subject to greater losses and, for iron, the peak rigidity is $\sim4\,{\rm EV}$. The corresponding energy is $\sim100\,{\rm EeV}$. Richer clusters can accelerate heavy nuclei to the highest energies measured, $\gtrsim200\,{\rm EeV}$. }
\label{fig:casspectrum}
\end{figure}
 
\section{Intergalactic Accretion Shocks}\label{sec:class}
We now apply these ideas to a model of a specific IAS, that associated with the Virgo cluster. In this section, we measure $u$ in units of $u_{\rm shock}=1000\,{\rm km}\,{\rm s}^{-1}$, $r$ in units of $r_{\rm shock}=2\,{\rm Mpc}$ and $\rho$ in units of $\rho_{\rm shock}=10^{-29}\,{\rm g\,cm}^{-3}$, adopting convenient estimates based upon the observations. This gives $L_{\rm shock}\sim2\times10^{45}\,{\rm erg\,s}^{-1}$ . This gives $R_{\rm diff}\sim10\,{\rm EV}$. For the interactions of protons with the CMB, the radiative loss rate can be approximated by $\dot q_{\rm rad}=-0.05R^{0.5}-5\times10^{-5}R^{2.2}\,{\rm Gyr}^{-1}$, for $2\lesssim R\lesssim200$. This yields a more stringent estimate $R_{\rm loss}\sim5\,{\rm EV}$.

A full treatment of this problem will include the gas, cosmic-ray and magnetic contributions to the time-dependent evolution, including the shock. It should also describe the downstream flow which affects the boundary condition.  Here we report on a simpler approach that seeks a stationary solution that matches the observations with plausible magnetic and particle energy densities. We suppose that the gas speed in scaled units is $u=1.18r^{-.5}-0.18r$, which has the Galaxy recede from Virgo at $u(8.5)\sim-1.1$. In addition, the shock is supposed strong with compression ratio 4 and the contribution of the high energy particles to the upstream pressure and energy density is subdominant. The domain of the particle disribution is taken to be the rectangle formed by $r=r_{\rm shock}=1$, $q=q_{\rm min}=-3$, $r=r_{\rm Galaxy}=8.5$, $q=q_{\rm max}=3$. At $q_{\rm min}$, we impose the test particle solution for $N$ with $F_r=0$ and normalization such that the proton energy density is below the kinetic energy density. $N$ vanishes at $q_{\rm max}$ and there is an outflow boundary condition for $r=r_{\rm Galaxy}$. As can be seen from Fig.~\ref{fig:casspectrum}, our prescription for $D(r,q)$ can lead to a spectrum of escaping particles with $R_{\rm max}\sim7\,{\rm EV}$, and accelerates a significant luminosity extending up to $\sim10\,{\rm EV}$. 

The proton rigidities achievable under this scenario are insufficient to account for the highest energies reported. However, DSA depends upon rigidity, and heavier nuclides can be accelerated to much higher energy. A proper treatment of this problem mandates a nuclear reaction network. Here, we just illustrate what could be happening by considering an exaggerated injection of iron nuclei which propagate in the turbulence generated by the protons with ${\dot q}_{\rm rad,Fe}=-0.07R^{2.2}\,{\rm Gyr}^{-1}$. The produces $R_{\rm max}\sim2\,{\rm EV}$, slightly less than those associated with protons. The associated energy is $E_{\rm max}\sim50\,{\rm EeV}$. More powerful shocks within the cosmic ray horizon can accelerate heavy nuclei to somewhat great energy, as observed. 

While it appears that rich clusters can accelerate the highest energy cosmic rays observed, we need less extreme accelerators to fill in the gap between the UHECR and galaxy-generated intergalactic cosmic rays. For these, we turn to the FAS. The same principles can be used as for cluster shocks with the important differences that filaments are essentially cylindrical, as opposed to spherical, that they are typically weaker with modest Mach numbers and that we may be much more interested in the cosmic ray population transmitted downstream. This is because the Galaxy may be located inside just such a filament.  Radiative loss is generally unimportant here. As the spectrum is much steeper, these cosmic rays should be mostly protons below the ankle in the spectrum.

\section{Galactic Wind Termination Shocks}\label{sec:GWTS}
Observations of the secondary to primary ratio of Galactic cosmic rays strongly indicate that they escape the Galactic disk in $\lesssim10\,{\rm Myr}$. Much of the hot gas from the supernova remnants that create them cannot cool and should likewise escape. The interstellar medium is an open system and these winds carry off mass, angular momentum and energy and help supply the circumgalactic medium with a population of $\sim1-10^6\,{\rm GeV}$ cosmic rays. More powerful galaxies and their active nuclei do likewise. These intergalactic cosmic rays are the dominant input for re-acceleration by IAS.

However, the escape velocity from the Galaxy's dark matter potential is several times the circular velocity and powering the wind presents a challenge. The best way to meet this interpretative challenge is to posit the presence of an MHD wind. In this case, gas only has to have sufficient specific enthalpy to be levitated to a modest height where it can pass through an Alfv\'en critical point and magnetic stress can take over to propel the combined hot gas plus cosmic ray fluid to radii $\sim200\,{\rm kpc}$. At some point, a Galactic wind must pass through a termination shock. Although this is definitely an MHD shock and somewhat different in character, it is expected to be an effective cosmic ray accelerator. Again, the shock is likely to be roughly spherical but the upstream flow is diverging, not converging, giving a negative contribution to $\dot q$. Against this, cosmic rays do not escape the upstream flow and can achieve modest, incremental energy gain from quite different parts of the shock as the mean free path at the highest energies may be an appreciable fraction of the radius. The contribution of the particles transmitted downstream fromall galaxies is plausibly consistent with the input cosmic ray spectrum invoked in Sec.~\ref{sec:class}.

\section{Supernova Remnants}\label{sec:SNR}
We take the next inferential step by applying the bootstrap approach to supernova remnant shocks. Specifically, we adopt the same basic prescription for the diffusion coefficient as we used for IAS and assume that most of the observed spectrum is dominated by the cosmic rays that escaped upstream. Of course, the accelerators are expected to be far more diverse with hot stellar wind shocks and supershocks formed by OB associations making significant contributions. This interpretation is quite different from the original motivation for DSA which presumed a single power law transmitted downstream. We now know that the measured spectra are far more structured and this could reflect an idiosyncratic local history.  An important part of traditional cosmic ray acceleration schemes is injection. In the DSA context these are the suprathermal particles that are chosen for acceleration to much higher energies. We have implicitly proposed that injection is unimportant in the termination and cluster shocks. The same could be true for supernova remnants if sufficient MeV cosmic rays are produced by stellar winds in the hot interstellar medium, prolonging their ionization loss times.

\section{Observational Implications}\label{sec:obi}
This hierarchical model for explaining most of the 36 octave, observed cosmic ray spectrum adopting one basic acceleration scheme, DSA plus bootstrap, is bold, simple and, probably, quite refutable. Despite it being the final stage of the process, we have highlighted UHECR production by IAS. The common features of DSA present at all stages are a quasi-spherical flow, the emphasis on the particles escaping upstream and the presence of strong MHD turbulence ahead of the shock by resonant cosmic rays. In the case of a IAS, however, the prediction is that the maximum energy  would not scale with rigidity, because of the limit imposed by the losses in the extragalactic photon backgrounds.  The bootstrap conjecture that there is a universal, local proportionality between the cosmic ray pressure and the energy density of the resonant waves is simple, efficient and easy to use. There is no question that this ansatz oversimplifies the problem  but it is easy to imagine a sequence of increasingly ambitious, high dynamic range simulations to explore it further.

The observational prediction is that there should be no signs of time-dependence, like what is expected from a transient model (see \citep{Globus:2023EA}). The secure dipole anisotropy (6.9~$\sigma$) and what becomes of less significant identifications should be consistent with the location of the Virgo cluster and richer clusters within the cosmic ray horizon after taking into account the Galactic magnetic field deflections. An intriguing possibility is that individual shocks be detectable through their radio synchrotron emission associated with much lower energy electrons and that the the totality of all such shocks accounts for the ARCADE 2 spectrum \citep[][and Simeon et al., these proceedings]{Simeon:2023UF}. The downstream transmission of cosmic rays, predicted on this basis, might come into conflict with upper limits on $\gamma$-ray emission from the outskirts of clusters from indirect searches for dark matter. Perhaps the most prescriptive implication of the Galactic wind stage is the prediction of a magnetic field that is strong and increasingly toroidal with radius (somewhat similar to what has been found in X-ray polarimetric observations of Pulsar Wind Nebulae and {\it in situ} measurements of the solar wind). Cosmic-ray propagation analyses for our Galaxy and infrared plus radio polarimetry of other galaxies are likely to confront these predictions. The identification of Galactic ``Pevatrons'' should help us understand if SNR are as efficient accelerators as suggested here. 

This work was supported by a grant from the Simons Foundation (00001470, RB, NG).
\bibliographystyle{JHEP}
\bibliography{references}{} 
\end{document}